%% file: astro2010_wp_kent_kaiser_rev2.tex
\documentclass[twocolumn,twoside,12pt]{article}

\usepackage{floatflt}
\usepackage{localepsfig}
\usepackage{graphics}
\usepackage{color}

\usepackage{pdfpart}
\usepackage{amssymb}
\usepackage{natbib}
\input{propdefs}

\begin{document}
\pagestyle{empty}
\title{\bf \Large Photometric Calibrations for 21$^{st}$ Century Science  \\
}
\renewcommand{\baselinestretch}{0.97}
\small \normalsize

\author{ \parbox{4.5in}{
\normalsize
Stephen Kent -- Fermi National Accelerator Laboratory, \\
~~~ MS 127, P.O. Box 500, Batavia, IL  60510, skent@fnal.gov \\
~ \\
Mary Elizabeth Kaiser -- The Johns Hopkins University, \\
~~~ Dept. of Physics \& Astronomy, 3400 North Charles St.,
Baltimore, MD 21218 kaiser@pha.jhu.edu  \\ 
~ \\
\renewcommand{\baselinestretch}{0.92}
\small \normalsize
Susana E.~Deustua -- Space Telescope Science Institute \\
J. Allyn Smith -- Austin Peay State University \\
Saul Adelman -- The Citadel \\
Sahar Allam -- Fermi National Accelerator Laboratory \\
Brian Baptista -- Indiana University \\
Ralph C.~Bohlin -- Space Telescope Science Institute \\ 
James L.~Clem -- Louisiana State University \\
Alex Conley -- University of Colorado \\
Jerry Edelstein -- Space Sciences Laboratory \\
Jay Elias -- National Optical Astronomy Observatory \\
Ian Glass -- South African Astronomical Observatory \\
Arne Henden -- Amateur Association Variable Star Observers \\
Steve Howell -- National Optical Astronomical Observatory \\
Randy A.~Kimble  -- Goddard Space Flight Center \\
Jeffrey W.~Kruk -- Johns Hopkins University  \\  
Michael Lampton -- Space Sciences Laboratory \\
Eugene A. Magnier -- Institute for Astronomy, U. of Hawaii \\
Stephan R.~McCandliss --  Johns Hopkins University \\  
Warren Moos -- Johns Hopkins University \\
Nick Mostek -- Lawrence Berkeley National Laboratory \\
Stuart Mufson -- Indiana University \\ 
Terry D.~Oswalt -- Florida Institute of Technology \\
Saul Perlmutter -- Lawrence Berkeley National Laboratory \\
Carlos Allende Prieto -- University College London \\
Bernard$\,$J.$\,$Rauscher -- Goddard Space Flight Center \\
Adam Riess -- Johns Hopkins University \\  
Abhijit Saha -- National Optical Astronomy Observatory \\
Mark Sullivan -- Oxford University \\
Nicholas Suntzeff -- Texas A\&M University \\
Alan Tokunaga -- Institute for Astronomy, U. of Hawaii \\
Douglas Tucker -- Fermi National Accelerator Laboratory \\
Robert Wing -- Ohio State University \\
Bruce Woodgate -- Goddard Space Flight Center \\
Edward L.~Wright -- University of California, Los Angeles \\
\\
{*Work supported by the U.S. Department of Energy under contract No.
DE-AC02-07CH11359.} \\
[.05in]}
}
\pagestyle{empty}
\maketitle
\normalsize
\onecolumn

%
\pagestyle{plain}
\pagenumbering{roman}
\setcounter{page}{1}
\setcounter{section}{-1}

\newpage

\clearpage
\setcounter{section}{0}
\setcounter{secnumdepth}{4}
\pagenumbering{arabic}
\setcounter{page}{1}
\twocolumn
\pagestyle{myheadings}
\markboth{ASTRO2010: Kent \& Kaiser}{\today}
\setlength{\baselineskip}{2.35ex}

\label{white1_kent}
\input{white1_kent}

\renewcommand{\baselinestretch}{0.96}
\small \normalsize

\label{astro2010_flux_stds_kaiser_revs3}
\input{astro2010_flux_stds_kaiser_revs3}

\renewcommand{\baselinestretch}{1.0}
\small \normalsize

\section{References}
\renewcommand{\baselinestretch}{0.75}
\small\normalsize
\small

\bibliography{references_astro2010}
%
%
\normalsize

\end{document}

%% file: propdefs.tex
\def\cc{\ifmmode {\rm cm}^{-3} \else cm$^{-3}$\fi}
\def\cl{\ifmmode {\rm cm}^{-2} \else cm$^{-2}$\fi}
\def\micron{\ifmmode \mu{\rm m} \else $\mu$m\fi}
\def\kms{\ifmmode {\rm km\,s}^{-1} \else km\,s$^{-1}$\fi}
\def\Hubble{\ifmmode {\rm km\,s}^{-1}\,{\rm Mpc}^{-1}
     \else km\,s$^{-1}$\,Mpc$^{-1}$\fi}
\def\ergsec{\ifmmode {\rm ergs\;s}^{-1} \else ergs s$^{-1}$\fi}
\def\ergscm{\ifmmode {\rm ergs\,s}^{-1}\,{\rm cm}^{-2}
       \else ergs\,s$^{-1}$\,cm$^{-2}$\fi}
\def\ergscmA{\ifmmode {\rm ergs cm}^{-2} {\rm s}^{-1} {\rm
\AA}^{-1}        \else ergs cm$^{-2}$ s$^{-1}$ \AA$^{-1}$\fi}
\def\ergscmHz{\ifmmode {\rm ergs cm}^{-2} {\rm s}^{-1} {\rm
Hz}^{-1}    \else ergs cm$^{-2}$ s$^{-1}$ Hz$^{-1}$\fi}



%% file: white1_kent.tex
\section{Introduction}

The answers to fundamental science questions in astrophysics, ranging from
the history of the expansion of the universe to the sizes of nearby
stars, hinge
on our ability to make precise measurements of diverse astronomical
objects.  As our knowledge of the underlying physics 
of objects improves  along with advances in detectors and instrumentation,
the limits on our 
capability to extract science from
measurements is set, not by our lack of understanding of the nature of
these objects, but rather by the most mundane of all issues: %
the precision with which we can calibrate observations in physical units.  

In principle, photometric calibration is 
 a solved problem -
laboratory reference standards such as blackbody furnaces achieve
precisions well in excess of those needed for astrophysics.
In practice, however, transferring the calibration from these laboratory
standards to astronomical objects of interest is far from trivial -
the transfer must reach outside the atmosphere, extend over 4$\pi$
steradians of sky, cover a wide range of wavelengths,
and span an enormous dynamic range in intensity.

Virtually all spectrophotometric observations today are
calibrated against one or more stellar reference sources, such as Vega, which
are themselves tied back to laboratory standards in a variety of ways.
This system's accuracy is not uniform.
Selected regions of the electromagnetic spectrum
are calibrated extremely well, but discontinuities of a few percent
still exist, {\it e.g.},  between the optical and infrared.  Independently,
model stellar atmospheres are used to calibrate the spectra of
selected white dwarf stars, e.g. the HST system, but the ultimate accuracy of
this system should
be verified against laboratory sources.
Our traditional standard star systems, while sufficient until now,
need to be improved
and extended in order to serve future astrophsyics experiments.

This white paper calls for a program to improve upon and expand the current
networks of spectrophotometrically calibrated stars to
provide precise calibration with an accuracy of equal to and better than  1\%
in the ultraviolet, visible and near-infrared portions of the spectrum,
with excellent sky coverage and large dynamic range. 

\vspace{-0.08in}
\section{Science Requiring Precision Calibration}
\vspace{-0.04in}

The following sections present four science investigations
that already are or soon
will be limited by the accuracy of photometric calibration.  This list is
not intended to be exhaustive.

\subsection{Expansion history of the Universe using Type Ia supernovae}

In 1998 we learned that the expansion of the universe is accelerating,
implying the
existence of a new component of the universe dubbed "dark energy".  
Precise measurement of the history of expansion and thus the properties of
dark energy is a major science goal of the next decade.  The Dark Energy
Task Force (DETF) \citep{Albrecht:2006}
has identified Type Ia supernovae as being one of four
principal methods for probing the expansion history. 

\begin{figure}[tbh]
\centerline{\includegraphics[width=2.19in,height=2.6in]
{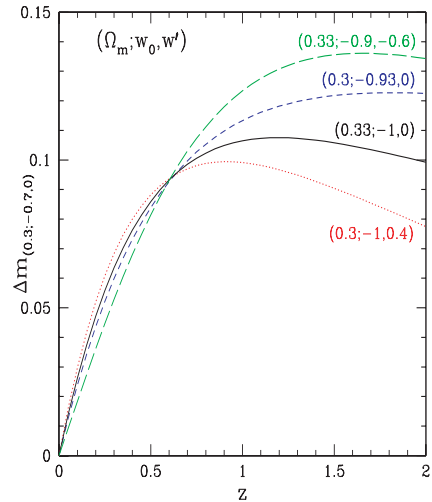}}
\vspace{-.14in}
{\renewcommand{\baselinestretch}{0.8}
\caption{\label{dmdz} {\small Differential magnitude-redshift diagram for dark 
energy models with $\Omega$, w$_0$, and w$^{\prime}=xw_a$. The 
difference between models is of order 0.02 magnitudes (or roughly 2\%). 
Models from \citealt{Huterer:2003}.}}}
\vspace{-.04in}
\end{figure}

Type Ia supernovae are thought to be ``standardizable candles'' -
from observations
of light curves and spectra, one can derive the luminosity of a supernova
that is the same on average with a scatter of $\approx$15\% for
a single object.
Cosmological and dark-energy parameters are
determined from the shape, not the absolute normalization, of the
Hubble brightness-redshift relationship. For each supernova, its
rest-frame B-band flux is plotted against its redshift, z. Since the
rest-frame B-band is seen in different bands at different redshifts,
the relative zero-points of all bands from 0.35 $\mu $m to 1.7 $\mu $m
must be cross-calibrated to trace the supernova from z = 0 to z =
1.7. 

Planned dedicated experiments, including
Pan-STARRS\footnote{http://pan-starrs.ifa.hawaii.edu/public/},
the Dark Energy Survey (DES) \citep{Abbott:2005},
the Large Synoptic Survey Telescope (LSST) \citep{Ivezic:2008},
and the Joint Dark Energy Mission\footnote{http://jdem.gsfc.nasa.gov/}
(JDEM)
and current and future observing programs 
using multipurpose facilities such as the 
Supernova
Legacy Survey on CFHT \citep{Astier:2006}, SN programs using 
Hubble Space Telescope \citep{Riess:2007}
and James Webb Space Telescope (JWST) \citep{Gardner:2006}
are or will be
focused on collecting accurate data for large numbers of supernovae,
eventually leading to a data set containing thousands of
objects ranging in redshift from 0 to 1.7.

\begin{figure}[tbh]
\centerline{\includegraphics[width=3in]
{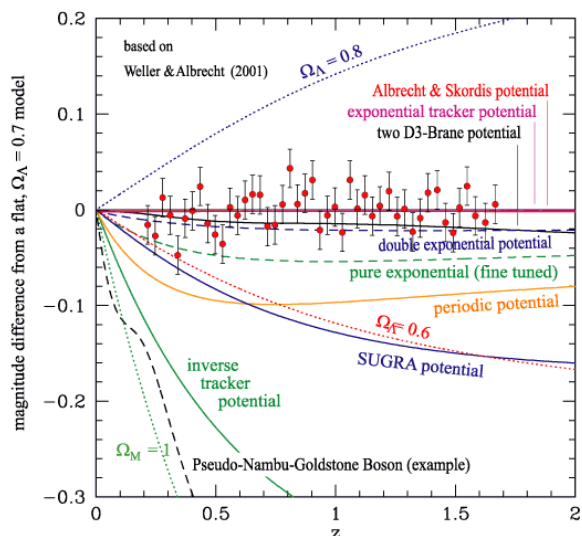}}
\vspace{-.14in}
{\renewcommand{\baselinestretch}{0.8}
\caption{\label{snplot} 
{\small Simulated SNIa data from one version of a JDEM mission compared
with predictions from a range of Dark Energy Models
(Derived from \citet{Weller:2001}).}}}
\vspace{-.04in}
\end{figure}

The power of using SNe$\,$Ia 
out to
z$\sim\,$1.7 for measuring the cosmological parameters is demonstrated
in Figure~\ref{snplot},
which compares the expected (simulated)
results of one version of a JDEM mission, to a range of possible  
dark energy models
\citep{Weller:2001} . This calculation is based on 2000 SNe Ia measured in the
range 0.1$\,\le\,$z$\,\le\,$1.7, plus 
 300 low-redshift SNe Ia from, e.g., the Nearby Supernova Factory
\citep{Aldering:2002}.  The simulated data have a statistical accuracy that
is capable of distinguishing models whose predictions differ by as little as
2\% over the full range of redshifts.

However, to make full use of the data, systematic errors must be comparable
to or smaller than the statistical errors.  
The NASA-DOE Joint Dark Energy Mission's Reference Mission specifies
that, over the fullwavelength range of 
0.35$\,<\,\lambda\,<\,1.7\,\mu$m, 
a  photometric uncertainty of 0.5\% per octave is required for the mission
to reach its target Figure of Merit.
Achieving this level of precision at the (faint) flux levels of the
redshifted SNe
requires a transfer of the absolute calibration from bright standard stars
to fainter calibration standard stars which can be directly observed by
the DE missions.

\subsection{Growth Of Structure}

\begin{figure}[tbh]
\centerline{\includegraphics[width=3in]
{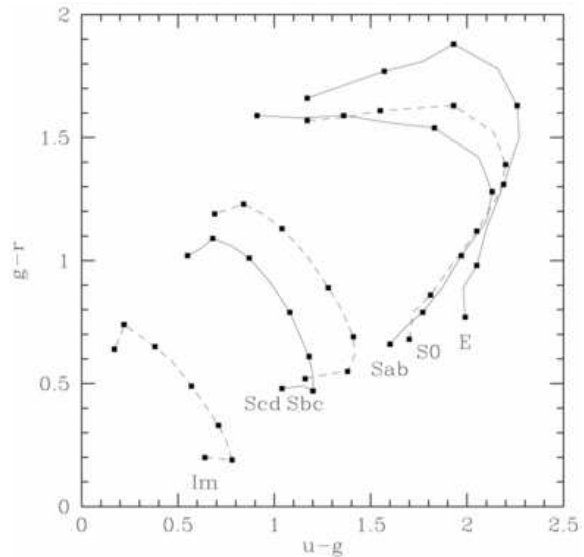}}
\vspace{-.14in}
{\renewcommand{\baselinestretch}{0.8}
\caption{\label{photoz} 
{\small A figure demonstrating the technique of photometric redshifts.
For each galaxy type, the dashed line joins together points that
mark the colors at a particular redshift.  From lower right to upper
left, the redshift increases from 0 to 0.6 (Eisenstein et al. 2001).}}}
\vspace{-.04in}
\end{figure}

A potentially powerful technique for measuring the
growth of structure in the universe is to use
gravitational weak lensing combined with
photometric redshifts of galaxies to study the statistical properties of
the mass distribution as a function of redshift.  The history of growth of
structure provides another approach to measuring the properties of Dark
Energy.  Current and future experiments that propose to collect data
for such studies include Pan-STARRS, DES, LSST, and JDEM.

A simplified description of this approach is as follows. One identifies
a set of galaxies at the same approximate redshift and measures the
distortions in the shapes of these galaxies induced
by gravitational lensing from the intervening mass distribution
(such as clustering).
A single set of galaxies measures the
properties of the integrated mass distribution along a line of sight.
By selecting a second set of galaxies at, {\it e.g.}, a higher redshift, and
measuring the changes in lensing-induced shapes relative to the first set,
one obtains information about mass structures in a slice of space between
the two sets of galaxies.  Thus, in a process  analogous to
tomography, one can build up a view of the mass structures and how
they change as a function of redshift.

A key necessity in this approach is the use of  multicolor, broadband
photometry of galaxies as a ``low-resolution spectrograph'' to estimate
redshifts  (Fig. \ref{photoz}).
Because the intrinsic spectral energy distribution of
any galaxy is not known a priori, one must rely on matching a set of
redshifted template spectra to the measured photometry of a galaxy and
utilizing a ``training set'' of galaxies with known redshifts to calibrate
the templates.  

Spectrophotometric calibrations are used to convert
the template spectra to predictions of galaxy magnitudes and colors. Ideally, 
the training set would span all of parameter space, but in reality there will
always be galaxies that can be measured photometrically but are too faint to
measure spectroscopically.  Accordingly, the LSST project has developed
a two-pronged approach to obtain photometric redshifts from
its multicolor data set \citep{Connolly:2006}, and 
 established a requirement on
spectrophotometric calibration of 1\% (1.5\% in the UV), with design goals 
that are twice as good\footnote{www.lsst.org/Science/docs/SRD.pdf}.

\subsection{Stellar Populations In Elliptical Galaxies}

Although elliptical and S0 galaxies are only a small fraction of all
galaxies, they are notable for having very similar stellar populations, as
reflected in their uniformity of colors. With the advent of large, multicolor
surveys using digital detectors, these objects can be identified over
a range in redshift and used for cosmological studies.  Thus, the red
galaxy spectroscopic sample in SDSS has been used to detect acoustic
baryon oscillations \citep{Eisenstein:2005}.  Additionally,
optical detection and measurement of
galaxy clusters has seen a resurgence of interest due to the ability to
identify galaxy clusters based on the ``red sequence'' of these types
of galaxies.  In low redshift clusters, the colors of  early-type
galaxies are remarkably uniform, showing a scatter of just 5\% in colors
such as SDSS $g-r$ and $r-i$ \citep{Koester:2007}.  The SpARCS
survey \citep{Wilson:2008} has shown that clusters with similar
galaxy content exist out to redshifts of at least 1.34.
Galaxy clusters will be detected and measured by nearly every current and
future imaging survey  conducted for weak lensing.  Galaxy
cluster counts have been identified as a third method for measuring
dark energy by the DETF.  

\begin{figure}[tbh]
\centerline{\includegraphics[width=3in]
{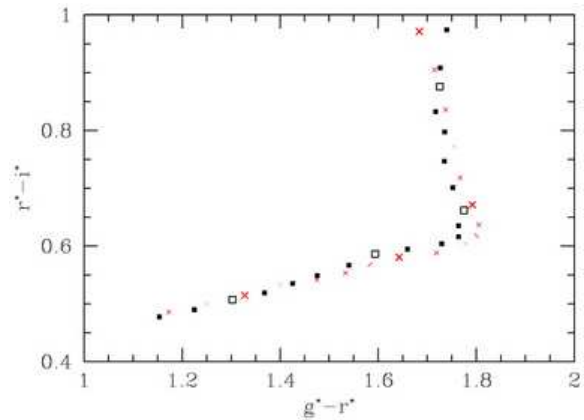}}
\vspace{-.14in}
{\renewcommand{\baselinestretch}{0.8}
\caption{\label{ellip} 
{\small Black squares: colors of elliptical galaxies as a function of redshift;
Red crosses: Passively evolving stellar population model
 (Eisenstein et al. 2001).}}}
\vspace{-.04in}
\end{figure}

Since a cluster
has anywhere from 10 to 100 members, the mean color of galaxies can be
measured with extremely high precision.  By comparing the colors over a
range in redshifts, it should be possible to make accurate models of the
stellar populations and infer their evolution over a significant fraction
of the age of the universe.  The limit on the accuracy of these models
will be set by the ability to self-consistently calibrate the galaxy
photometry over the optical and near-IR bands.  Conceivably one could take
advantage of data calibrated at better than 1\% accuracy.  Figure
\ref{ellip} demonstrates the precision with which elliptical galaxy
colors can be measured and compared with stellar synthesis models.

\begin{figure}[tbh]
\centerline{\includegraphics[width=3in]
{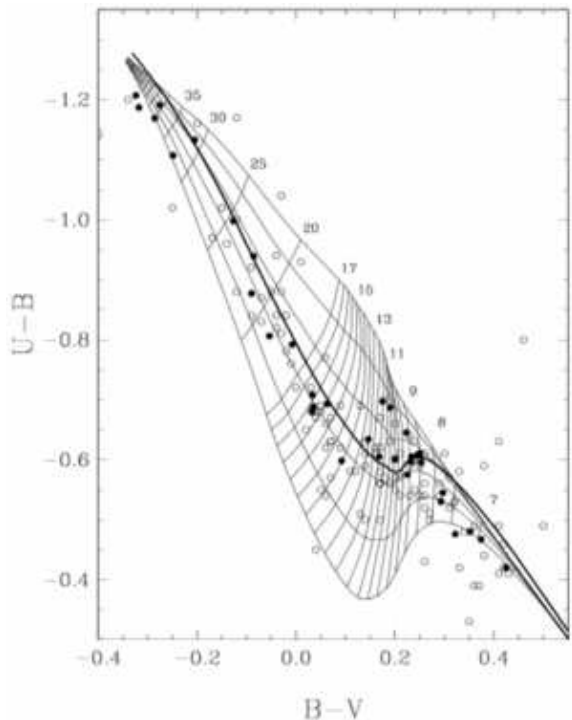}}
\vspace{-.14in}
{\renewcommand{\baselinestretch}{0.8}
\caption{\label{wd}
{\small Color-color diagram for DA white dwarfs.  Open and filled circles
are observations of stars with measured distances.  Solid lines are
predictions from a grid of models with constant gravity or constant
effective temperature \citep{Holberg:2006}.}}}
\vspace{-.04in}
\end{figure}

\subsection{Stellar Structure}

The fundamental parameters of stars,
including mass, radius, metallicity, and age, are inferred by matching
accurate models of stellar atmospheres to calibrated spectroscopic data and thus determining 
 effective temperature, surface  gravity, composition
 and, if necessary, interstellar reddening.  For stars with
relatively simple atmospheres such as hydrogen white dwarfs,  atmosphere
models are thought to be quite accurate and can be used to predict
photometric parameters (Fig. \ref{wd}) and, in combination with
stellar interior models, the radii and absolute luminosities as
well.  By combining these data with photometric measurements, it is possible
to predict distances.  A comparison of these predictions with measured
trignometric parallaxes for those stars with such measurements shows
excellent agreement \citep{Holberg:2008}.
If calibrations can be improved to the level of 1\% and with more stars
(such as will be measured with GAIA),
it will be possible to make meaningful tests of 3-D
spherical models,
derive masses directly, and make more quantitative tests of
evolutionary models.

%% file: astro2010_flux_stds_kaiser_revs3.tex
\vspace{-.10in}
\small
\section{Flux Calibration \& Standardization}
\normalsize
\vspace{-1mm}

Ultimately, observed astrophysical fluxes must be converted to
physical units.  Three of the most common methods of determining the
absolute fluxes are through comparison to standard stars (e.g.~solar
analog stars), stellar atmosphere models, and certified laboratory
standards. But, the existing precision of each of these methods is
inadequate to meet the requirements of the science described in the
previous section.

\vspace{-2mm}
\subsection{Solar Analog Stars}
\vspace{-1mm} 

Use of solar analog stars as a standard source relies upon the star
having the same intrinsic SED as the sun. Unfortunately, no star is a
true solar analog.  Even G-type stars with the most-closely matching
visible spectra can differ 
by a few percent.  In addition, uncertainties in the solar SED itself
are 2-3\% \citep{Thuillier:2003}.

\vspace{-2mm}
\subsection{Stellar Atmosphere Models}
\vspace{-1mm}

UV and visible astrophysical fluxes are often normalized to an
absolute flux using a set of hot, white dwarfs (WDs) whose models are tied
to Vega's absolute flux at 5500 Angstroms, as determined through
direct comparison to a black body reference.

Stellar atmosphere models are currently the
preferred method for calibrating stellar fluxes due to the agreement
between the models and the observations as well as the increased resolution
of both the models and the data.  Use of these pure hydrogen WD stars
simplifies the computation and improves the precision by eliminating
one of the most difficult steps in atmospheric modeling - that of
including the blanketing from the plethora of metal lines.

To obtain the absolute flux and its uncertainty for an unreddened
WD, medium-resolution high S/N ($>\,$50) observations of the Balmer lines are
fit to model hydrogen line profiles to determine the effective
temperature, the gravity, and the associated uncertainties
\citep[e.g.,][]{Finley:1997}. Then, the best-fit model and the
models at the extremes of the uncertainty in T$_{\rm{eff}}$ and
log$\,$g determine the nominal flux and uncertainty in the shape
of the flux distribution.  These model fluxes are normalized to
V-band Landolt photometry.

\begin{figure}[tbh]
\vspace{-5mm}
\centerline{\includegraphics[width=3.4in,height=3.4in]{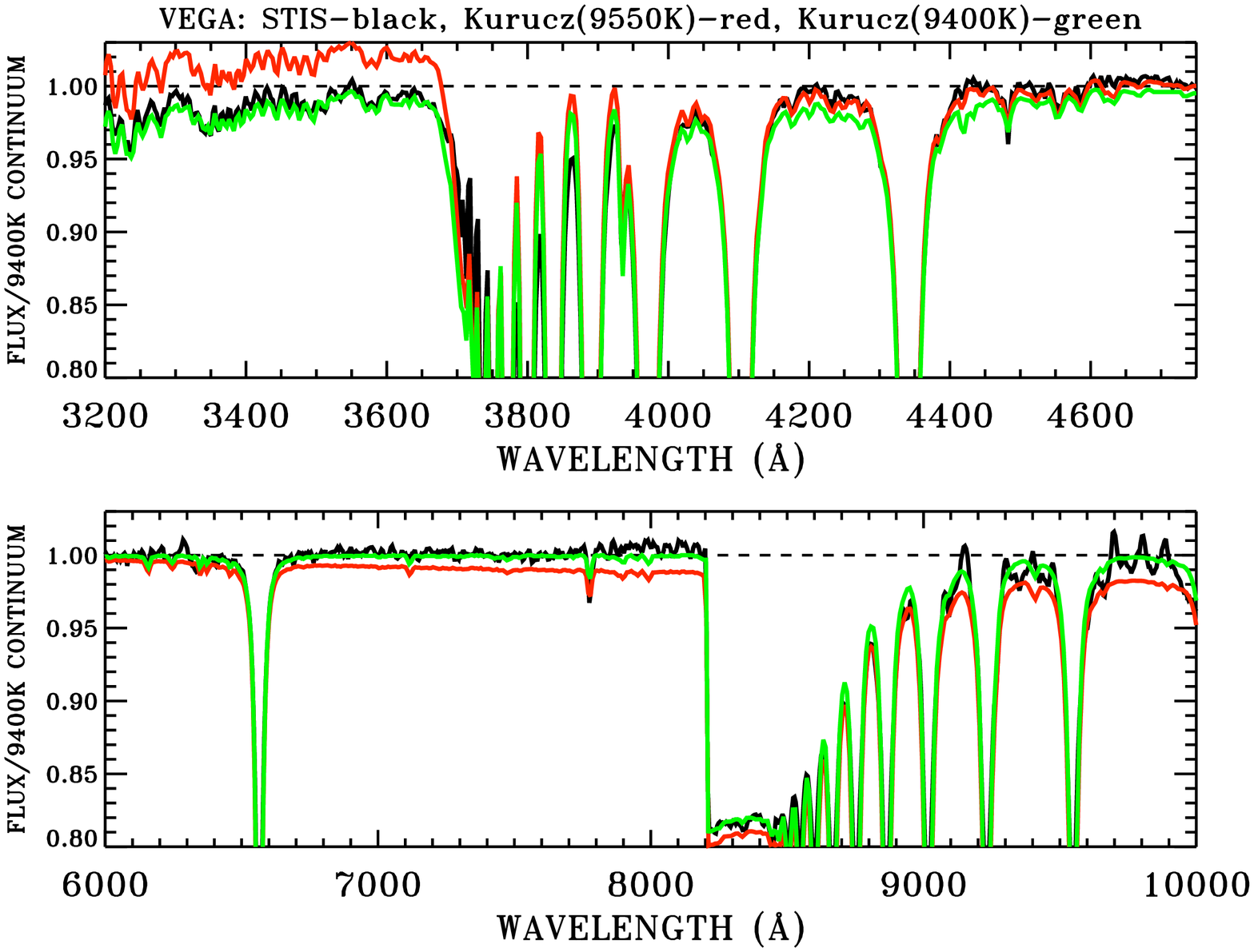}}
\vspace{-10mm}
{\renewcommand{\baselinestretch}{0.75}
\caption{\label{vega} {\small Uncertainties in the absolute flux for Vega: 
HST/STIS observations (black line: \citealt{Bohlin:2007, Bohlin:2004a}), 
the Kurucz stellar model with T$_{eff}$=9400$\,$K
(\textcolor{green}{green}), and the Kurucz stellar model at 9550$\,$K
(\textcolor{red}{red}) are compared.  The observations exhibit better agreement
with the cooler model at the longer and shorter wavelengths.  The
hotter model agrees better with the measured flux by $\sim\,$1\% at
4200--4700$\,$\AA.}}}
\vspace{-6mm}
\end{figure}

The three primary WD standards of {\it{HST}} CALSPEC network are
internally consistent to an uncertainty level of 0.5\% in the
visible with localized deviations from models rising to $\sim$1\%
over the 4200$-$4700$\,$\AA\ spectral range, and a $\pm\,$1\%
uncertainty in the NIR (1$-$2$\,\mu$m) (Fig.~\ref{vega};
\citealt{Bohlin:2007}). 
Current uncertainties in the extensive NIR
(1.0$\,<\,\lambda\,<\,$1.7$\,\mu$m) network of standard stars are
$\sim$2\% (e.g.~\citealt{Cohen:1992a, Cohen:1992b, Cohen:2003b, Cohen:2007}).

Any systematic modeling errors that equally affect the shape of the
flux distributions of all three WD stars cannot be ruled out
and would make the actual error larger.  
Differences between the continua of the LTE and NLTE models place a
lower limit of ~2\% on the uncertainty in the 0.35$-$1.7~$\mu$m range
for these standards.

In the NIR, astrophysical fluxes are often normalized to A-star
models, where the accuracy of the best A-star models rivals that of
the pure hydrogen WD models.  Absolute photometry of Vega is used to
normalize the SEDs of these stars to an absolute flux scale.
\citet{Rieke:2008} tested the agreement of IR standard star
calibrations and models based on direct absolute measurements of A0V
stars versus the sun and examined the impact of extrapolating the IR
data into the visible.  The data were found to be consistent,
permitting flux calibrations with an accuracy of $\sim$2\% between 1
and 25$\,\mu$m.


\vspace{-2mm}
\subsection{Certified Laboratory Standards}
\vspace{-1mm}


Photometry of Vega has been absolutely calibrated
against terrestrial observations of certified laboratory standards
(e.g.\ tungsten strip lamps, melting point black bodies)
to provide the normalization for the network of stellar models and
templates that are used as practical absolute standards.  These
absolute calibrations to standard sources were difficult and subject
to large systematic uncertainties due primarily to the large and
variable atmospheric opacity.

Discrepancies of $\,>\,$10\% in Vega's flux exist at 0.9$-$1$\,\mu$m,
whereas the measurements from 0.5$-$0.8$\,\mu$m agree to
$\sim\,$1\% (\citealt{Bohlin:2004a, Hayes:1985}). Beyond 1$\,\mu$m, windows of low water vapor
absorption have been used for absolute photometry
(e.g.~\citealt{Selby:1983, Mountain:1985}).



Currently, the uncertainty in the standard star flux calibration
network relative to the fundamental laboratory standards exceeds 1\%.


{\bf {Certified Detectors:}} The calibration precision of
photodetectors has greatly improved since early pioneering measurements
(e.g.~\citealt{Oke:1970, Hayes:1975}).  Current NIST
$\sim$2$\,\sigma$ uncertainties in the absolute responsivity of
standard detectors are $\sim\,$0.2\% for Si
photodiodes and 0.5\% for NIR photodiodes \citep{Larason:2008}.  This
increased precision in the photodetector calibration, ease of use,
and repeatability, now make standard detectors the calibrator of choice.



\vspace{-2mm}
\subsection{Extension to Standard Star Networks}
\vspace{-2mm}

The basic techniques and methodologies for extending one fundamental
standard candle to a network of stellar standards are well
established.  This extensive network of stellar standards is
fundamentally tied to the sun or to Vega, e.g~ SDSS successfully
established a network of standard stars spanning the visible range
from 0.3$-$1.0~$\mu$m \citep{Smith:2002} with absolute fluxes based on
Vega using BD+17$^{\circ}$4708 as an intermediate (V=9.5 mag) transfer
standard \citep{Fukugita:1996}. Even the \citet{Cohen:1992a} absolute
standard models of Sirius are tied to Vega as the underlying standard.


Vega is far too bright to be observed directly by the current class of
4-m telescopes and even with most 2-m telescopes using
state-of-the-art instruments. Its use as a standard is further
complicated by its protoplanetary disk which contributes to IR flux
measurements.  In addition, as a pole-on rotator its surface
temperature and gravity vary dramatically from the pole to equator (e.g.~\citet{Aufdenberg:2006}).
This introduces complexity into accurately and
precisely representing its flux with robust stellar atmosphere
models. Furthermore, uncertainties in atmospheric corrections have resulted in
wavelength dependent uncertainties in Vega's intrinsic flux.
Thus, Vega is not suitable as a modern astrophysical flux
standard.


NIST standards have been transferred to observations of other stars,
but the level of uncertainty in the flux measurements have precluded
their widespread use (e.g.  HZ43 and G191B2B: $\sim$4\% precision,
\citet{Kruk:1997}).  An exception to this was the Midcourse Space
Experiment (MSX) which observed eight standard stars, including Vega,
in the infrared and directly tied these observations to inflight
measurements of emmisive reference spheres \citep{Price:2004a}. These
measurements resulted in corrections (Sirius: 1\%) and caveats (Vega:
flux excess). These MSX observations were
limited to bright, typically K$\,$III and M$\,$III, stars in six
selected NIR/IR bandpasses.  Thus, the need for a sample of absolutely
calibrated astrophysical standards spanning a broad dynamic range in
flux and wavelength (UV through NIR) persists.


Current astrophysical problems need a precise (better than
1\%) network of astrophysical flux standards spanning a wide dynamic
range. This enables scientists to take advantage of the capabilities
of current and future telescopes and the instruments that were
developed to address pressing scientific questions.
New, direct measurements of standard stars tied
directly to fundamental NIST standards are required. 

\vspace{-2mm}
\subsection{Current Status \& Future Prospects}
\vspace{-1mm}

Although the relative photometry of objects in a single CCD
exposure can be better than 1\%, this level of precision is not
achieved for the relative fluxes of sources in different fields of
view.  \citet{Stubbs:2006} reviewed systematic uncertainties that plague
ground-based observations, discussed the challenges associated with
characterization of atmospheric transmission and the removal of
instrument artifacts, and presented a method for achieving photometry
with fractional uncertainties.  
Using precisely calibrated photodiode detectors in concert with a
wavelength tunable laser illumination source, \citet{Stubbs:2007a}
demonstrated the success of their methodology in measuring the
instrument transmission and established the capability of standard
detectors as a fundamental metrology to achieve precise and accurate
photometry.

Other programs are also making concerted efforts to
characterize instrument performance (e.g~ASTRA \citet{Adelman:2007}),
however, the need to monitor and correct for atmospheric transmission
on short timescales persists. One approach 
(e.g.~Pan-STARRS, LSST) uses a dedicated telescope to monitor the
atmosphere  throughout the night to enable
corrections for science observations at the neighboring facility.

Direct, absolute calibrations of stellar fluxes measured above the
Earth's atmosphere are also being pursued. A recently
approved sub-orbital program, ACCESS: Absolute Color Calibration
Experiment for Standard Stars (\citealt{Kaiser:2008, Kaiser:2007}),
will transfer NIST absolute detector standards to additional standard
stars with better than 1\% precision over the $\sim$3500\AA\ $-$
1.7$\mu$m bandpass at a spectral resoloving power of
$\sim$500. However, due to the limited time above atmosphere for
rocket flights, these measurements will be limited to a few stars
brighter than $\sim$10$^{th}$ magnitude.

The scientific impact of a standard star network based on the absolute
calibration of stars too bright to be observed with the premier
telescopes needs to be addressed. A modern calibration program should
extend direct flux measurements to fainter sources, encompass a broad
spectral range (UV through the IR), ensure robust results through the
support of independant calibration programs, and provide technology
support to execute these programs.

In conclusion, we stress the need for a calibration program that
supports the science of the 21st century.